\documentclass[a4paper,12pt]{article}
\usepackage{inputenc}
\usepackage{amssymb,amsmath}
\usepackage[english]{babel}
\usepackage{epsfig,graphicx,color}

\newcommand{\be}{\begin{equation}\label}
\newcommand{\ee}{\end{equation}}

\newcommand{\bib}{\bibitem}

\begin{document}

\begin{center}

{\bf \large Urusovskii's Geometry, Algebrodynamics\\   
and Universal Quantum-like Kinematics in Complex Space}

\vskip2mm

{\bf Vladimir V. Kassandrov}

\vskip2mm

{(\it Institute of Gravitation and Cosmology, \\ 
Peoples' Friendship University, Moscow, Russia)}

\end{center}

\vskip2mm

\section{The Minkowski space-time model: does it really represent 
the ultimate physical geometry?}

Nowadays, the Minkowski geometry necessarily constitutes the 
framework  of any realistic physical geometry including 
its curved version in General Relativity, multidimensional schemes etc. 

However, the true dimension and structure of real space (and time) is still 
unclear, and there is still no answer to the old question: ``Why {\it  
``visible''} physical space is 3-dimensional?'' And what in fact 
can be said on the origin and properties of {\it Time}? 

From general physical viewpoint, the Minkowski space $\bf M$, together with 
the Lorentz group as its group of symmetry, possesses a number of 
remarkable and well-known properties (causal 
structure, universal velocity of propagation of interactions etc.), 
is well grounded in theoretical (symmetries of Maxwell and Lorentz equations) 
and purely experimental respects. 

On the other hand, there are some known drawbacks of the STR paradigm 
(absence of the global evolutionary parameter, of phase relations, 
of time irreversibility etc.). And the most grievous fact is that 
$\bf M$ {\it does not originate from first principles}, its 
mathematical structure is rather cumbersome and not distinguished 
in any aspect. Besides, assumption  of $\bf M$ as physical 
geometry {\it does not offer any key to comprehend the dynamics of fields and 
particles or the quantum properties of matter!} 

Thus, one has to return back to foundations and seek for an 
internally exceptional structure of (extended) physical space-time which 
by itself could predetermine the properties of matter!

In this note, we briefly present two approaches to the 
establishment of the extended space-time structure and, 
combining them, demonstrate a purely classical, geometrical 
explanation of the quantum interference phenomena. 

First approach (section 2) is the ``6D treatment of Special Relativity'' 
proposed by Igor A. Urusovskii in 1996-2011 (see~\cite{UrusPIRT} and references 
within). Key notions of his approach are: 1) universal 
light-like velocity of {\it any} particle in the whole 6D space, and 
2) universal internal rotation of particles in the 3D space 
orthogonal to the physical 3D one. 

Second approach (section 3) is my own {\it ``algebrodynamical''} one (see 
references, say, in~\cite{PIRT03,PIRT09}) in which one is also brought to a 6D geometry of 
special type. Key constituents arising naturally in the framework 
of algebrodynamics are (sections 4-6) 1) the ensemble of identical copies 
({\it``duplicons''}) 
of one and the same particle and, 2) {\it ``dimerous''} structure of 
electron which turns to be detectable only at the instants of 
merging of some two duplicons. 

Finally, rather simple and natural combination of the above two approaches 
(section 7) immediately follows in a clear classical picture of the canonical 
{\it two-slit interference experiment} (section 8). Concluding remarks are 
presented in section 9. 

\section{On the 6D geometry of Urusovskii}

Approach presented by I.A. Urusovskii in his original papers~\cite{UrusClass} 
is based in fact on the following two axioms (conjectures): 

1) {\it universal light-like motion} of all of the  pre-elements of matter 
(say, electrons) in the extended (3+3) physical space (so that {\it none particle 
is ever at rest}!). The principle was earlier discussed by F. Klein, Yu.B. Rumer, 
R.O. di Bartini et al and can either substitute or supplement the principle of 
{\it universal velocity of interaction propagation} in Special Relativity. 

2) {\it universal periodical regular rotation} of particles along a 
curcumference of {\it Compton radius} $R=\hbar/Mc$  in the additional 3D-space
(orthogonal to the ordinary physical 3-space). This conjecture closely relates to 
the L. de Broglie's hypothesis on {\it internal frequency} $\nu$ 
(internal clocks) of elementary particles, $\hbar \nu = Mc^2$, to the 
conception of {\it Zitterbewegung} of E. Shr\"odinger etc. 

Consequently, Urusovskii naturally assumed that {\it proper time} of a particle (being, 
say, at rest in ``observable'' 3-space and rotating, therefore, with constant 
speed $c$ in orthogonal 3-space) {\it is proportional to the number of turns 
(or to phase increment) in its  rotation in the internal space}. 

Then for a particle moving uniformly with velocity $v$, velocity of rotation 
$u$ is obtained from the principal relation $v^2+u^2=c^2$ and determined 
by the familiar Einstein's factor, 
\be{decel}
u=c\sqrt{1-v^2/c^2}. 
\ee
Thus, one comes to the canonical effect of {\it time deceleration} for a moving 
particle, even {\it apart of any specification of a transformation symmetry 
group}. 

As to the latter, Urusovskii himself assumed it to be the Lorentz group 
since ``the number of turns = the proper time interval'' should be invariant, 
the same for any observer.    
  
Moreover, according to Urusovskii, {\it all principal relations of STR 
including the dynamical ones, follow in fact from purely Newtonian physics in 
the entire 6D space, under projection onto the ``observable'' 3D space}! 
For more details of Urusovskii's approach see his original papers~\cite{UrusClass}; 
on further development of the theory see, e.g.,~\cite{UrusPIRT} and references 
therein. 

Let us remark now that the first postulate, on the universal light-like 
velocity, is obviously equivalent to the following relation for the 
increments of particle's  coordinates $\{{\bf X,Y}\}$ in the physical ${\bf X}$ and 
internal ${\bf Y}$ spaces: 
\be{diff}
dx_1^2+dx_2^2+dx_3^2+dy_1^2+dy_2^2+dy_3^2= c^2 dt^2.
\ee
We see thus that {\it one deals in fact with a Euclidean 6D space}, whereas 
the {\it global universal time should be considered as the metric on this space}, 
not as a coordinate! In this respect, the Urusovskii's viewpoint on his theory 
as on a ``6D treatment of STR'' is, at least, not evident or uniquely possible.  
Note also that recently a number of attempts were undertaken {\it to  
reformulate STR on the basis of the 4D Euclidean space} (see, e.g.,~\cite
{Montanus,Pestov}).

As to the second postulate on regular internal rotation (in a 2D plane), 
it can be implemented for a particle at rest ($dx_1=dx_2=dx_3=0$) 
if one requires, say, in addition to (\ref{diff}):                                                                
\be{rotat}
dy_3=0,~~~dy_1^2+dy_2^2=ds^2=R^2 d\varphi^2 = c^2 dt^2 ,  
\ee 
where the proper time increment $ds$ turns to be the metric on {\it internal} 
space, whereas $d\varphi$ is the corresponding phase increment.

From further results obtained by I. Urusovskii in the framework of his 6D approach 
one can distinguish, say, 1) the {\it hydrogen spectrum} with account of its 
fine structure, 2) the {\it quark model} of nucleons, 3) a new treatment of 
gravitation (as a projection of {\it cosmological force} retaining particles 
in a Compton-order vicinity of ${\bf R}^3$), 4) a novel treatment of the 
{\it Universe expansion} (resolving a number of paradoxes in standand 
cosmology) etc. 

Despite of these results (which one can give credence to or not), it is 
especially important that Urusovskii's approach allows {\it to link the 
space-time geometry with phase relations} (what could probably lead to a   
{\it geometrization of quantum theory}) and to directly deduce universal 
{\it kinematics} of particles (and, to some extent, their {\it dynamics} as 
well) from pure {\it geometry}!

However, in many aspects the Urusovskii's approach is certainly insufficient. 
Indeed, any general substantiation of the starting postulates is absent, 
the problem of {\it mass spectrum} is not resolved, structure and kinematics 
of {\it photons} (which, in the considered framework, should be treated as 
non-rotating in ${\bf Y}$) remain quite unclear etc. Evidently, a more general and 
mathematically profound approach is needed. In the following section we 
briefly present the algebrodynamical scheme that can aspire to such a role.

\section{Main principles and results of algebrodynamics}

In the {\it algebrodynamics}~\cite{AD,Qanalys} one assumes the existence of a  
({\it exeptional} in its internal properties) {\it space-time algebra} (STA). 
The STA predetermines both the geometry of physical space-time (via the {\it 
automorphism group} acting as a Klein's symmetry group of metric), and the 
particle-field dynamics (via the {\it analyticity conditions} for 
functions-fields over the STA acting as field equations). 
Particles themselves may be confronted with {\it singularities} 
of functions-fields. 

In the role of STA it had been considered the {\it algebra of complexified 
quaternions} (biquaternion algebra $\mathbb B$), equivalent to the matrix 
$Mat(2,\mathbb C)$-algebra.
The $\mathbb B$-analyticity  conditions -- generalized {\it Cauchy-Riemann 
equations} (GCRE) -- {\it turn to be nonlinear as a consequence of noncommutativity}. 
Thus, one deals in fact with fields with (self-)interaction!

Some consequences of GCRE are: emergence of gauge and spinor (twistor) 
structures, identical satisfaction of Maxwell and Yang-Mills free equations 
(for secondary derivative fields), {\it self-quantization} of electric charge
~\cite{Sing} etc.

Note also that the geometry induced by the structure of $\mathbb B$-algebra is 
4D in complex numbers = 8D in reals. Whether one (rather artificially!) 
reduces the coordinate space to real ${\bf M}$, the whole theory becomes 
{\it Lorentz invariant}! 

A novel concept of physical Time had been also elaborated in which time 
manifests itself as a evolution parameter related to the 
condition of {\it local preservation of the primordial $\mathbb B$-field} 
(= principal spinor or twistor field in other equivalent representations). 

Local evolution turns to be a {\it light-like transfer} of the primary 
$\mathbb B$-field. On this way, one comes to the concept of the {\it Flow of 
PreLight = Flow of Time}~\cite{Pavlov,Levich}. 
Geometrically, particles manifest themselves as {\it caustics of the PreLight 
Flow} so that {\it whole matter is of a light-like nature}!

\section{Complex-quaternionic geometry of space-time}

It is noteworthy, however, that none algebra is known which could give rise 
precisely to ${\bf M}$. Nonetheless, the considered 4D (in $\mathbb C$) = 8D  
(in $\mathbb R$) ${\mathbb B}$-induced geometry has the $SO(3,\mathbb C)$ 
symmetry group which is 6-parametric and isomorphic to the Lorentz group! 

On the other hand, ${\bf M}$ does not contain any {\it phase-like} component, 
whereas the ${\mathbb B}$-induced geometry naturally decomposes into a 
Minkowski-type {\it macro-geometry} (defined by the modulus parts of complex 
coordinates) and {\it micro-geometry} of a ``fiber'', related to their phase 
parts~\cite{Mink05} and responsible, probably, for the wave properties of 
matter. 

Actual ${\mathbb B}$-geometry of (extended) space-time turns to be 6D and 
resembles the geometry proposed by I. Urusovskii. It is dynamical in origin 
being closely related to universal kinematics of particles in the primordial 
complex ${\mathbb B}$-space. Such kinematics can't be realized in real Minkowski 
space and results in a rather unexpected picture of Physical World. 
One can thus assume that {\it we really live in the complex space-time}!

\section{Complex null cone and the concept of duplicons}.
  
Complex space-time geometry allows for realization of the {\it one-electron Universe} 
conjecture of Wheeler-Feynmann.  On the {\it real} space-time background this is in 
fact impossible. Indeed, in real ${\bf M}$, according to the {\it retardation 
equation (= equation of the light cone)}, one has only one unique object in the 
past that determines the field at a given point  (for $v<c$; for tachions -- 
more than one, see, e.g.,~\cite{Bolot}).

On the contrary, in complex space $Z=\{z_0,z_1,z_2,z_3\}$ equation of the 
{\it complex null cone} 
\be{complexcone}
(z_1-{\hat z}_1(\tau))^2+(z_2-{\hat z}_2(\tau))^2+(z_3-{\hat z}_3(\tau))^2=
(z_0-\tau)^2, ~~~\tau \in {\mathbb C} 
\ee
can have a great number of roots (dependently on the form of 
${\mathbb C}$-trajectory ${\hat Z}(\tau)$), 
in accord, say, with the ``principal theorem of algebra'' (for  
polynomial functions ${\hat Z}(\tau)$). 

\begin{figure}
\centering
\includegraphics[width=10cm,height=5cm]{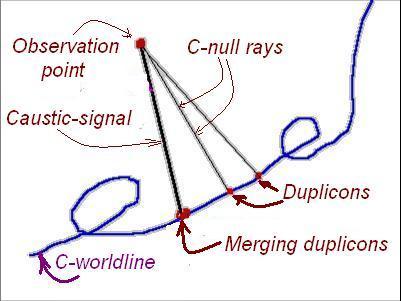}
\caption{\small Duplicons and caustic formation.}
\end{figure}

Therefore, an observer ``living in a ${\mathbb C}$-space'' instantaneously 
receives the field signals from a lot of copies locating at the points on the 
same ``complex worldline'' (Fig.1). Thus, one actually deals with an ensemble of 
identical particle-like formations -- ``duplicons''~\cite{PIRT05,YadPhys}. 
Points of observation and location of a duplicon are connected by a $\mathbb C$-null 
straight line -- ``ray'' (quite similar to the case of real ${\bf M}$ where 
one deals with local light cones). 
These lines densely filling ${\mathbb C}$-space form a 
fundamental {\it null congruence} for which the ${\mathbb C}$-worldline itself 
play the role of its {\it  focal line}~\cite{PIRT05}. 

\section{Caustics as lightlike signals and ``dimerous electron''}

At some {\it discrete} instants (for a corresponding position of an observer 
in the ${\mathbb C}$-space) a {\it merging} of a pair of duplicons does take 
place (Fig.1). This relates to {\it multiple roots} of the ${\mathbb C}$-null cone 
equation (\ref{complexcone}). Again, such an event is in principle impossible 
in real ${\bf M}$.

At these moments strong {\it amplification} of electromagnetic and other fields 
emerges, along a  null straight line connecting points of observation and 
of duplicons' merging. Geometrically, these lines represent the {\it caustic 
locus} of the fundamental null congruence~\cite{PIRT05}. 
Physically, one observes a  {\it lightlike pulse -- a signal} coming from 
the points of duplicons' mergings. 

Thus, in the framework of a purely classical electrodynamics but contrary to the 
case of real ${\bf M}$ background, in complex world there occure  
{\it quantum-like discrete radiation processes}! 

One obtains therefore a {\it self-consistent relativistic dynamics of an 
ensemble of identical and causally connected ``particles''}. 
It is evident, however, that (contrary to the Wheeler-Feynmann paradigm) 
individual duplicon {\it in principle} can't be considered as a primary 
matter pre-element: only a pair of these can be detected, and only at discrete 
instants of their merging. We are forced thus to surmise that ``electron'' 
not only {\it consists} of two ``halves'' - duplicons, but does not in fact 
{\it exist} as a unique formation, except at discrete instants of 
merging - radiation - detection acts.

Conjecture on {\it electrons as dimerons}~\cite{PIRT09,YadPhys} directly 
correlates with modern notions on ``fractal charges'' and ``electron bubbles''~\cite{Bykov}, 
``light-shining-through-wall'' effects~\cite{tunnel} etc. It allows for 
{\it alternative explanation of the wave  properties of matter} (section 8).

\section{Links between complex and Urusovskii's geometries}

In full analogy with the case of real ${\bf M}$, the primary field and the 
caustic structure both {\it reproduce themselves along the null complex straight 
lines} --  elements of the $\mathbb C$-null cone~\cite{PIRT05}. 
For their infinitesimals one has
\be{inf}
dz_0^2+dz_1^2+dz_2^2+dz_3^2=0.
\ee
Decomposing $dz_\mu = dx_\mu + i dy_\mu$ and separating real and imaginary 
parts in (\ref{inf}), one obtains a system of {\it two quadrics} in 
${\bf R}^8$, 
\be{quadric}
dx_0^2+dx_1^2+ dx_2^2+ dx_3^2-dy_0^2-dy_1^2-dy_2^2-dy_3^2 =0, 
\ee
\be{quadric2}
dx_0 dy_0 +dx_1 dy_1 +dx_2 dy_2 +dx_3 dy_3 =0 .  
\ee
First constraint (\ref{quadric}) defines in fact {\it two 4D Euclidean 
(sub)spaces}; these are moreover mutually orthogonal as a result of the second 
equation (\ref{quadric2}). We can then define the {\it time increment} $cdt$ as  
{\it common metric} on the both subspaces, so that relation (\ref{quadric}) 
takes the form: 
\be{timemetric}
dx_0^2+dx_1^2+ dx_2^2+ dx_3^2 = dy_0^2+ dy_1^2+ dy_2^2+ dy_3^2 := c^2 dt^2.  
\ee
It is natural to assume that fundamental kinematics of matter pre-elements 
follows just from the condition of preservation of the primary field and 
caustic structure and 
satisfies therefore condition (\ref{inf}) or, equivalently, (\ref{quadric2})  
and (\ref{timemetric}). According to these, {\it universal ``eternal'' motion 
of matter pre-elements occures in both mutually orthogonal spaces $\bf X$ and  
$\bf Y$, with one and the same, constant in modulus fundamental 
``velocity of light''} $c$. 

Thus, complex geometry leads by itself to satisfaction of the first postulate 
of the Urusovskii's scheme. Specifically, for a particle which is at rest in 
the  ``physical'' 3D subspace of $\bf X$, one obtains making use of 
(\ref{quadric2}) and (\ref{timemetric}):
\be{3Drest}
dx_1=dx_2=dx_3=0,~~\rightarrow~~dy_0 =0,~~~ 
dx_0^2= c^2 dt^2 = dy_1^2+ dy_2^2+ dy_3^2, 
\ee  
so that it necessarily moves with the speed of light in the 3D subspace 
of $\bf Y$. 

The geometry allows moreover for exact {\it rotating configurations} 
considered by Urusovskii. To obtain these, one can require, say, in addition 
to (\ref{3Drest}):
\be{rotatconf}
dy_3 =0,~~dy_1^2 +dy_2^2 = R^2 d\varphi^2~~(= c^2 dt^2 = dx_0^2),
\ee
We see therefore that the increment of zeroth coordinate $dx_0$ (generally, 
of $dz_0 = dx_0+i dy_0$), invariant under transformations from the symmetry group 
$SO(3,\mathbb C)$ of ${\mathbb B}$-algebra, appears in the role of a 
{\it proper time interval}~\cite{Mink05}, in full 
correspondence with the Urusovskii's picture. It is clear, however, that 
much more general configurations are admissible, namely (for a particle 
resting in 3D subspace ${\bf X}$), an arbitrary light-like motion in the 
orthogonal 3D subspace ${\bf Y}$. 

\section{Phase of internal revolution and quantum interference}

For Urusovskii's configurations applied to a duplicon its {\it phase change} 
$d\varphi$ along a trajectory is proportianal to the increment of proper time $ds$, 
\be{phase}
d\varphi = ds/R, 
\ee
where $R$ is the radius of internal rotation which, according to Urusovskii, is 
equal to the Compton length of the duplicon. However, for correspondence with 
quantum theory (see below), we shall assume for this radius  
\be{radius}
R=\hbar/2Mc,    
\ee
where $M$ is taken to be the mass of the particle (electron) associated with 
duplicons.

\begin{figure}
\centering
\includegraphics[width=10cm,height=7cm]{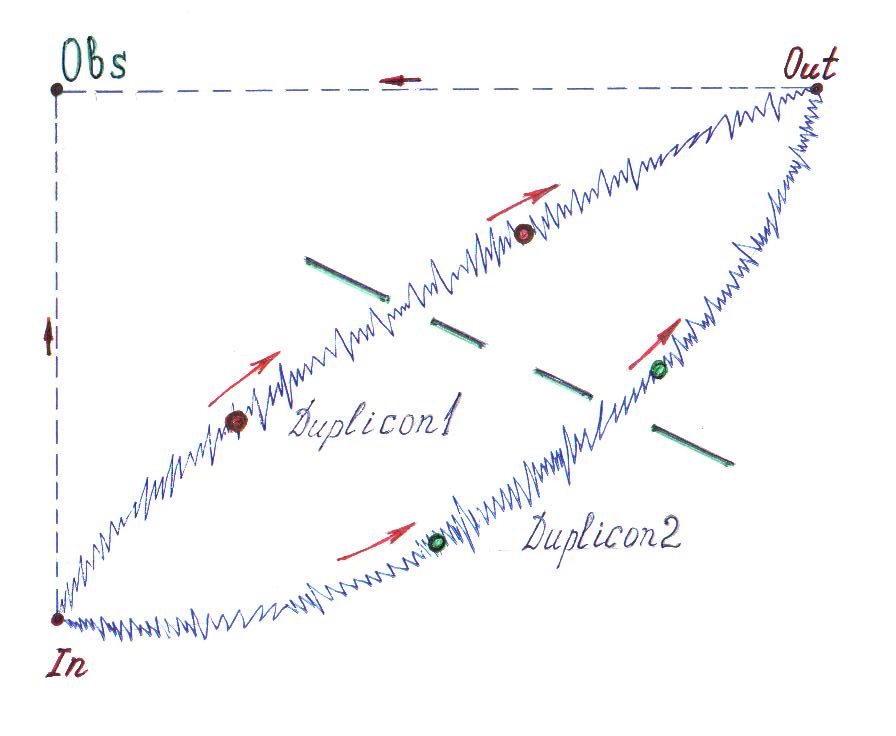}
\caption{\small Successive mergings of duplicons in a two-slit experiment.}
\end{figure}

In~\cite{PIRT09,YadPhys}  it had been already demonstrated that, together with 
the ``dimerous electron'' conjecture, fundamental relation (\ref{phase}) 
along with (\ref{radius}) leads to a purely classical explanation of the 
canonical two-slit experiment. Specifically, one can consider this as two 
successive mergings of a pair of duplicons at space-time points where they can 
be detected by an observer (Fig.2). Since these mergings occure in fact in the 
extended complex space-time, the coordinates of two duplicons should coincide 
whereas their phases could differ to $2\pi N,~~N\in {\mathbb Z}$. Thus, the phase 
shift $\Delta \varphi$ for a pair of duplicons between their successive 
mergings should be 
\be{phaselag}
\Delta \varphi = \Delta \int d\varphi = 2\pi N, 
\ee
or, making use of (\ref{phase}) and (\ref{radius}), 
\be{interf}      
\frac{2Mc}{\hbar}\Delta \int ds = 2\pi N. 
\ee
This is {\it general relativistic invariant condition for two successive mergings 
of duplicons} which substitutes the canonical condition for 
interference maxima. Indeed, in the first order of non-relativistic 
approximation $v/c <<1$, making use of representation 
\be{approx}
ds=cdt\sqrt{1-v^2/c^2}\approx cdt - v dl/2c,
\ee
with $dl=v dt$ being the path length increment,  
one obtains instead of the principal condition (\ref{interf}):
\be{nonrelativ} 
\Delta \int \frac{dl}{\lambda(l)} = N, 
\ee    
where $\lambda(l)$ denotes the de Broglie's wavelength (generally, 
variable along a trajectory),
\be{wavelength}
\lambda(l):= h/Mv .
\ee

Obviously, equation (\ref{nonrelativ}) corresponds to the familiar 
condition for interference maxima (for constant velocity it takes the form 
$\Delta L = N\lambda$). However, general relativistic condition in 
the algebrodynamics is represented just by equation (\ref{interf}) which may be 
rewritten in the form
\be{interf2}
\Delta S = \frac{N}{2} \Lambda, 
\ee
where $S$ is the length of a 4D worldline in real ${\bf M}$ for one of 
duplicons (between two successive merging events), and $\Lambda=h/Mc$.

\begin{figure}
\centering
\includegraphics[width=10cm,height=5cm]{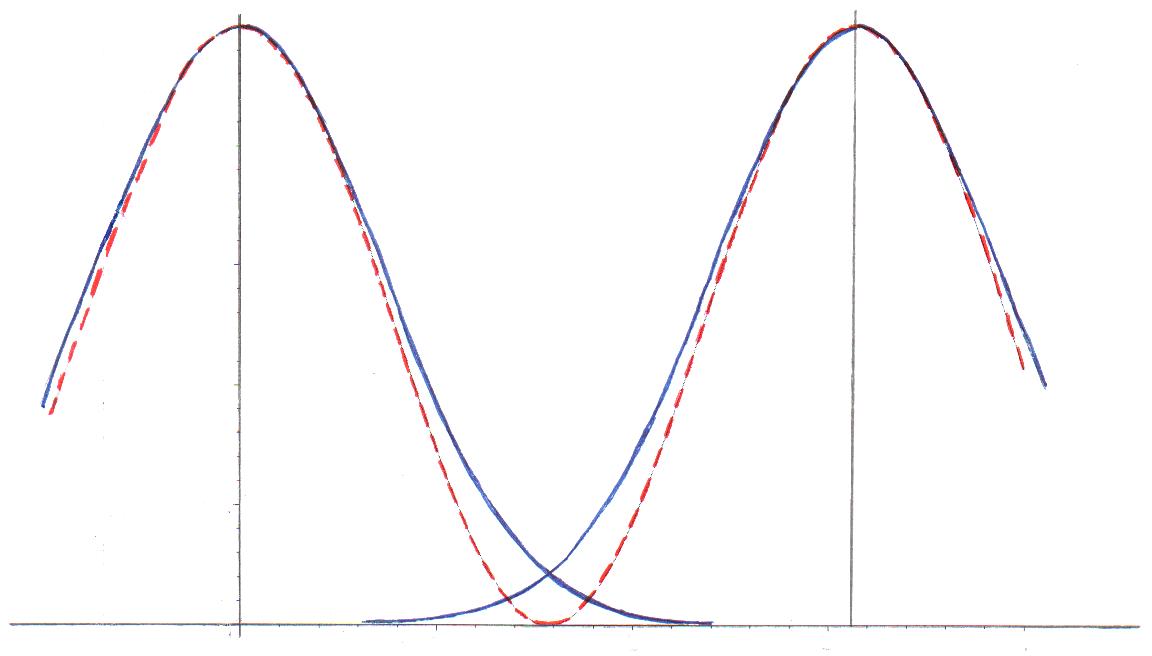}
\caption{\small QM vs AD distributions of probabilities in a two-slit 
experiment.}
\end{figure}

Thus, in complex algebrodynamics, in the idealized situation, one deals with 
a $\delta$-shaped probability distribution for interference pattern since 
duplicons can merge and be thus detected only at some discrete space points 
specified by the condition (\ref{interf2}). However~\cite{PIRT09}, due to statistical 
errors in a real experiment (during preparation, detection processes etc.) 
the probability distribution smears and takes, near a single maximum, 
a Gaussian-like form (Fig.3, solid line)
\be{gauss}
\sim \exp(-\Delta x^2/\Lambda^2),
\ee
while the canonical wave-like distribution is of a form (Fig.3, dashed line)
\be{canon}
\sim \cos^2(\Delta x/\Lambda).
\ee
In vicinity of a maximum these coincide up to the 3-d order derivative.  
Nonetheless, the distinction (see Fig.3) could be revealed  
in a suitable quantum interference experiment.  

\section{Concluding remarks}

We have demonstrated that quantum interference phenomena can be explained  in 
a purely geometrical manner whether one adopts two independent fundamental 
conjectures  dealing with 1) universal kinematics of matter pre-elements 
(rotation with velocity of light in the additional space) and, 2) dimerous 
nature of electrons and, probably, other elementary particles 
(detectable at some particular instants of merging of their constituents -- 
duplicons). 

We have shown that these two conjectures are compatible with each other but, 
unfortunately, first one does not follow directly from the fundamental 
algebraic dynamics in complex space, so that Urusovskii's universal kinematics 
remains up to now an intuitive though very attractive hypothesis. 

Nonetheless, let us mention that, actually, there exist much more rigid 
restrictions on kinematics of pre-particles in the complex-quaternionic space 
than those revealed in section 7. These are based on twistor 
structure of the primary ${\mathbb B}$-field and make the Urusovskii's 
conjecture even more realistic.

\end{document}